\documentclass[pdflatex,sn-mathphys]{sn-jnl}

\graphicspath{{fig/}}

\jyear{2022}%

\theoremstyle{thmstyleone}%
\theoremstyle{thmstyletwo}%

\theoremstyle{thmstylethree}%

\begin{document}

\title[Diffusion-driven frictional aging in silicon carbide]{Diffusion-driven frictional aging in silicon carbide}

\author*[1]{\fnm{Even Marius} \sur{Nordhagen}}\email{evenmn@mn.uio.no}

\author[1]{\fnm{Henrik} \sur{Andersen Sveinsson}}\email{h.a.sveinsson@fys.uio.no}

\author[1]{\fnm{Anders} \sur{Malthe-Sørenssen}}\email{malthe@fys.uio.no}

\affil[1]{\orgdiv{The Njord Center and Department of Physics}, \orgname{University of Oslo}, \orgaddress{\street{Sem Sælands vei 24}, \postcode{NO-0316}, \state{Oslo}, \country{Norway}}}



\abstract{
    Friction is the force resisting relative motion of objects.
    The force depends on material properties, loading conditions and external factors such as temperature and humidity, but also contact aging has been identified as a primary factor.
    Several aging mechanisms have been proposed, including increased ``contact quantity'' due to plastic or elastic creep and enhanced ``contact quality'' due to formation of strong interfacial bonds.
    While proposed mechanisms for frictional aging have been dependent upon the presence of a normal force, this factor is not a fundamental prerequisite for the occurrence of aging.
    In light of this, we present a novel demonstration of a substantial frictional aging effect within a cubic silicon carbide system, even when a normal force is entirely absent.
    Our observations indicate that the time-evolution of the frictional aging effect follows a logarithmic behavior, which is a pattern that has been previously observed in numerous other materials.
    To explain this behavior, we provide a derivation that is rooted in basic statistical mechanics, demonstrating that surface diffusion, a phenomenon that serves to minimize surface energy in the interface region, can account for the observed behavior.
    Upon application of a normal force, the friction force is enhanced owing to the presence of plastic creep.
    Although aging resulting from plastic and elastic creep is widely recognized and incorporated into most friction laws, diffusion-driven aging has received comparatively less attention.
    The ultimate objective is to develop or redesign friction laws by incorporating the microscopic behavior, with the potential to enhance their effectiveness.
}



\maketitle

\section{Introduction}\label{sec:introduction}
Frictional aging is a phenomenon in which the strength of the interface between two contacting surfaces changes over time under static loading conditions.
This idea was first suggested by Ernst Rabinowich in a pioneering study from 1951, in which he proposed that frictional aging is responsible for the dissimilarity between static and dynamic friction \cite{rabinowicz1951}. Additionally, it has been proposed that frictional aging is responsible for the velocity-weakening effect observed in dynamic friction \cite{dieterich1978}.
A deeper understanding of frictional aging is important as it affects the behavior of many mechanical systems, such as brakes, clutches and bearings.
Furthermore, frictional aging plays a role in many natural phenomena, such as earthquakes \cite{scholz2019a}, landslides \cite{handwerger2016} and glacier flow \cite{lipovsky2017}, so studying it can lead to a better understanding of these events and how to mitigate their impacts.
Moreover, viewed from a physics perspective, friction and frictional aging is a subject of considerable interest \cite{krylov2014}.

Two possible mechanisms of processes causing frictional aging have been proposed \cite{dieterich1972}. 
One is the increase in the actual area of adhesive contacts between surfaces, also known as ``contact quantity''. The other mechanism involves an increase in the strength of these contacts, referred to as ``contact quality''. Both of these mechanisms have been observed in experiments and are believed to play a significant role in frictional aging. 
A study by Dieterich and Kilgore demonstrated how the contact area increases with contact age due to creep \cite{dieterich1994}, while Li et al. showed that stronger interfacial bonds can form without a corresponding increase in contact area \cite{li2011}.
Simultaneous occurrence of the two mechanisms is also plausible \cite{li2022}.
Determining which mechanism that causes frictional aging in individual cases is among several aspects that are not fully understood.
The time-dependent nature of both increased contact quantity and enhanced contact quality, both of which exhibit a logarithmic behavior, can make it challenging to ascertain the correct mechanism. 
This difficulty in identifying the underlying mechanism makes it currently impossible or extremely challenging to predict the frictional behavior based purely on material properties.
As of now, the friction models in use are phenomenological, and their parameters must be derived from experiments rather than being pure material constants.

In order to enhance our understanding of frictional aging, it is crucial to employ reliable methods.
Atomistic simulations, for instance, have been proposed as a promising method to bridge the gap between microscale friction behavior and macroscale laws \cite{dong2013,spikes2018}.
Atomistic simulations can offer valuable information that may be difficult to acquire through real-world experiments.
Specifically, since the positions, velocities, and forces of all particles are known at each time step, the spatial and temporal resolution is only restricted by the length of a time step.
Molecular dynamics simulations have already demonstrated their remarkable capability as friction simulators \cite{zhang1997,zhang2001,mo2009,liu2012,li2014}, and have provided insights that are currently unattainable in the laboratory with the available technology.

\begin{figure}
    \centering   
    \includegraphics[width=10cm]{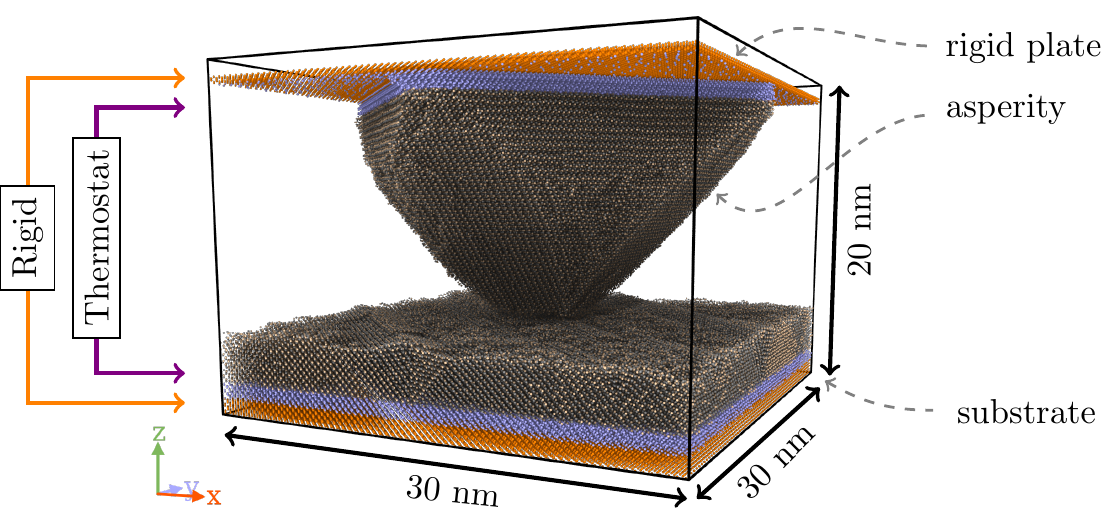}
    \caption{Illustration of the system immediately after the asperity and the substrate are brought in contact. During the simulations the lower part of the substrate and the plate that the asperity is attached to are held rigid (orange layers). The upper part of the asperity and the layer above the rigid layer of the substrate are controlled by a thermostat (blue layers). System dimensions are marked with double-headed arrows.
    }
    \label{fig:regions}
\end{figure}

To gain insight into frictional aging at the microscopic level, we examine a silicon carbide asperity on a silicon carbide substrate.
Recrystallization, a process that occurs in systems far from equilibrium, causes the actual contact area between the two objects to grow in this system.
Recrystallization is a typically slow process that necessitates longer time scales than are available in atomistic simulations.
For example, quartz crystals develop naturally over many years under high pressure and temperature, while diamond formation may take hundreds of millions of years \cite{stachel2015}.
However, silicon carbide has a relatively low activation energy that results in significant self-diffusion at intermediate temperatures \cite{ghoshtagore1966,hong1979,sveinsson2020}.
This feature enables atomistic simulations of processes that are impractical for most other materials.
In this study, we investigate the frictional aging of silicon carbide interfaces using molecular dynamics simulations at temperatures where surface diffusion is substantial.
Silicon carbide friction experiments have been conducted both in the laboratory \cite{zhao2014} and through computer simulations \cite{piroozan2019}, and they are often motivated by industrial applications \cite{lafon-placette2015}.
To our knowledge, experiments have not yet been conducted in temperature regimes where recrystallization resulting from surface diffusion is significant.


\begin{figure*}
    \centering
    \includegraphics[width=11.8cm]{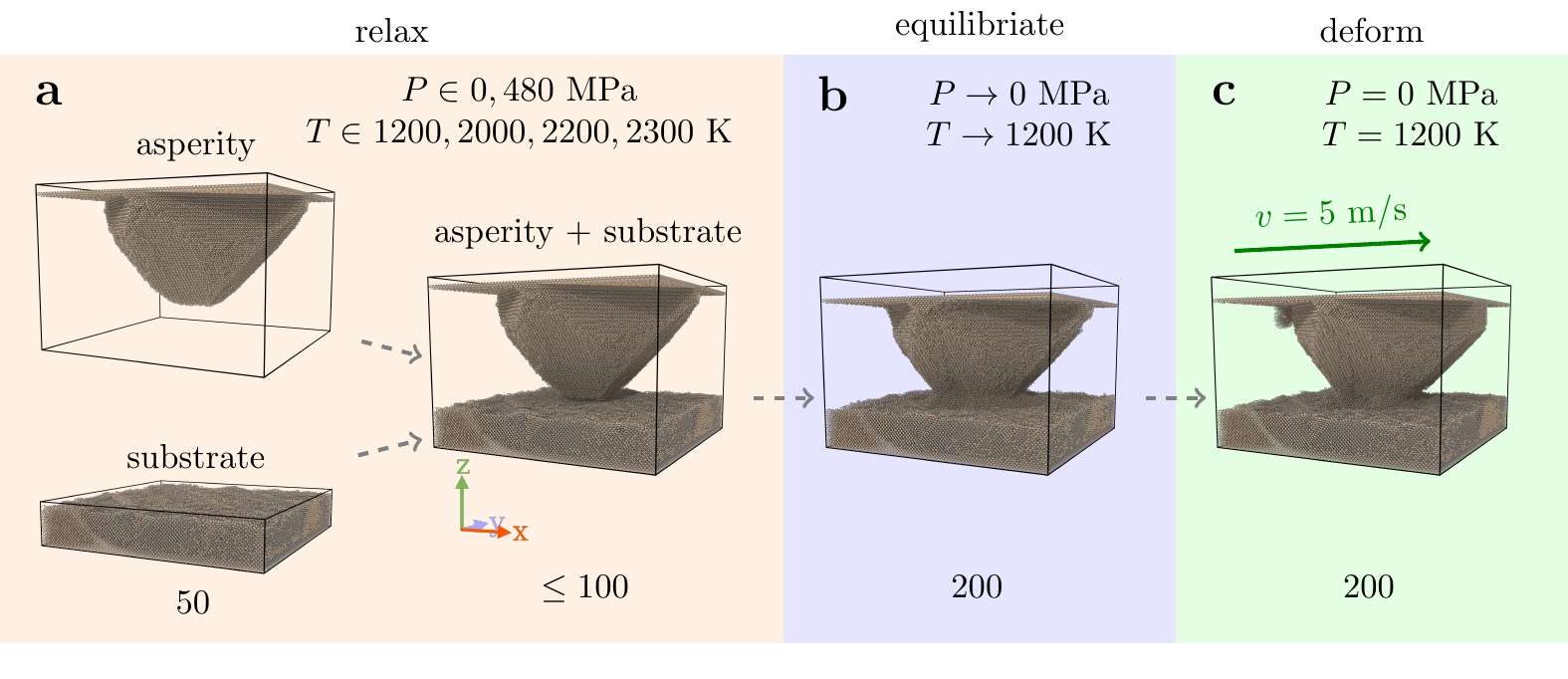}
    \caption{An illustration of the three simulation steps. In (a) an asperity and a substrate are relaxed separately for 50 ns, at a temperature $T$. Thereafter, they are brought in contact and relaxed for up to 100 ns under a pressure $P$. In (b) the system is equilibriated by gradually reducing the pressure to zero and the temperature to 1200 K, for 200 ps. In (c) the system is deformed by applying a constant velocity on the upper part of the asperity for 200 ps.}
    \label{fig:steps}
\end{figure*}

\section{Simulations}\label{sec:method}
We perform molecular dynamics simulations to study aging and frictional properties of a silicon carbide nano-asperity on a silicon carbide substrate, as illustrated in figure \ref{fig:regions}.
The nano-asperity is prepared by carving out a rectangular octahedron from a cubic silicon carbide (3C-SiC) crystal with distance 11.19 nm between the (110) planes and the center of the asperity.
Then, a rhombic dodecahedron with distance 11.70 nm between the (111) planes and the center of the asperity is carved out making the asperity shape close to the equilibrium shape of our model silicon carbide \cite{sveinsson2020}.
The asperity is attached to a rigid top plate that is used to impose a normal load and lateral motion on the asperity. The normal load is imposed by adding a force to the top plate particles, and the lateral motion is imposed by controlling the lateral position of the top plate.
The bottom 1 nm of the substrate is rigid and fixed in all directions to keep the substrate in place, and the 1 nm above the fixed layer is kept at a constant temperature using a Langevin thermostat.
The uppermost 1 nm of the asperity (just under the top plate) is also kept at constant temperature using a Langevin thermostat.
The non-rigid parts are integrated using Newton's 2nd law, to avoid disturbing the dynamics in the process region.
This is the \textit{de facto} standard setup of atomic friction simulations, and have been seen in many recent studies \cite{dong2013,li2014,schall2021}.
The overall system size is $30\times 30 \times 20$ nm, with 866,918 atoms, half of them silicon and half of them carbon. 

\begin{figure}
    \centering
    \includegraphics[width=11.8cm]{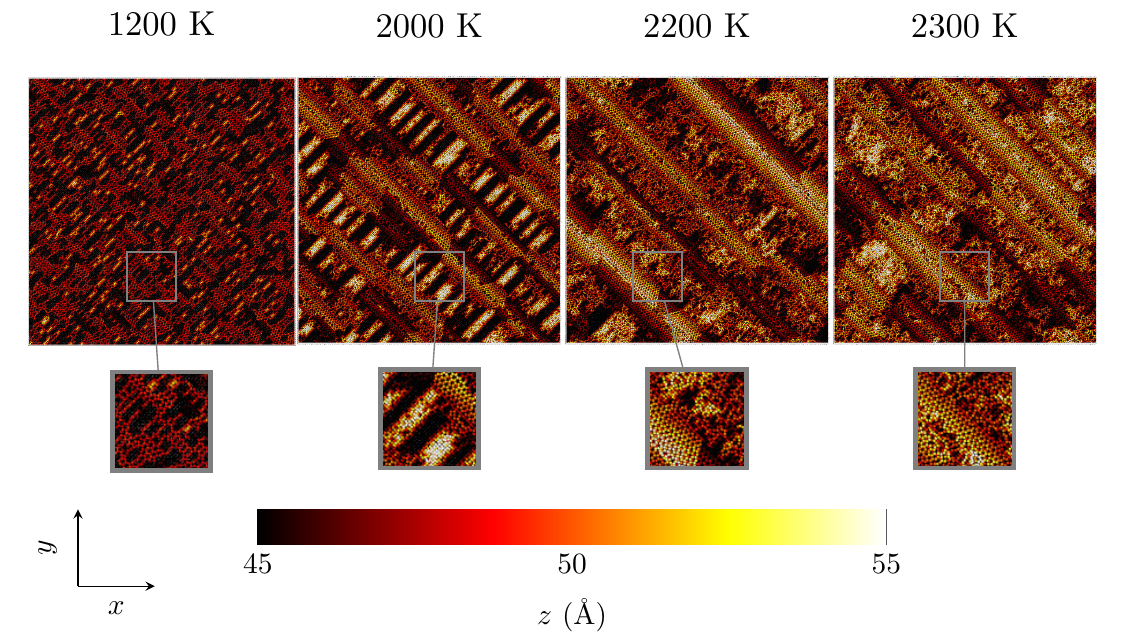}
    \caption{The figure shows the top view of the substrates following a 50 ns relaxation period at various temperatures. Magnified windows are included to highlight interesting regions of the substrates. The atoms are colored based on their $z$-position, where brighter colors indicate larger $z$-components.}
    \label{fig:lower}
\end{figure}

To assess the aging effect on static friction of the silicon carbide nano-asperity, we perform molecular dynamics simulations in the following three steps:
First, we relax an asperity and a substrate in separate simulation cells for 50 ns at a temperature in the range [2000, 2300] K to imitate a realistic system and ensure stable simulation outcomes. For the case with an applied normal force, we run an additional simulation at low temperature ($T=1200$ K). During this time the asperity relaxes to its equilibrium shape, and the substrate becomes rough. 
Figure \ref{fig:lower} displays substrates relaxed for 50 ns at some selected temperatures and is discussed below.
Thereafter, the asperity and substrate are brought into the same simulation box, separated by a distance of 4 Å, and relaxed. During this relaxation, the asperity is pulled towards the substrate due to a cohesive force originating from induced charge-dipole and dipole-dipole interactions. The substrate and the asperity are therefore spontaneously brought into contact. We then let the system evolve through time at a temperature in the range [2000, 2300] K and pressure in {0, 480} MPa and monitor the contact area (Figure \ref{fig:steps}a). The simulation is run for 20 ns (apart from a simulation at $T=2300$ K, $P=0$ MPa that was run for 70 ns), and during this time the contact area grows. We refer to the simulation time in this phase as the contact age. 

We extract system configurations at different contact ages. These systems are equilibriated by gradually decreasing the temperature to 1200 K, and the pressure to 0 MPa over a time period of 200 ps (Figure \ref{fig:steps}b).
Lastly, the static friction of the system is assessed by translating the rigid layer of the asperity at a constant velocity of 5 m/s in $x$-direction and measuring the shear force by which the system resists this translation (Figure \ref{fig:steps}c).

\subsection{Equilibrating the substrate}
Nano-ribbons are visible on the equilibrated substrates, as shown in figure \ref{fig:lower}.
Major nano-ribbons (ribbons traversing boundary to boundary) are present at all temperatures that are examined.
These ribbons are formed because the unstable (100) plane starts converting into a hexagonal ring structure, which is expected since cubic silicon carbide converts to hexagonal structure at approximately 2100 K \cite{ryan1968}.
Surprisingly, we observe the formation of hexagonal surface structure even at 1200 K.
Conversion from cubic to hexagonal structure starts on the surface since it requires expansion as the density of the hexagonal structure is smaller than that of the cubic structure.
At temperatures around 2000 K, we observe additional small ribbons between the major ribbons, which still have a cubic crystal structure.
We anticipate the equilibrium crystal structure to be reconstructed from Wulff construction.
The difference in temperature can be attributed to thermally activated diffusion, which becomes chaotic at high temperatures ($T>2100$ K).
The surface nano-ribbons are unlikely to significantly impact the static friction behavior since our analysis is focused exclusively on the static friction force and not the sliding friction.
Sliding friction will be accounted for in a future paper.


\section{Results and discussions} \label{sec:results}
\begin{figure}
    \centering
    \includegraphics[width=11.8cm]{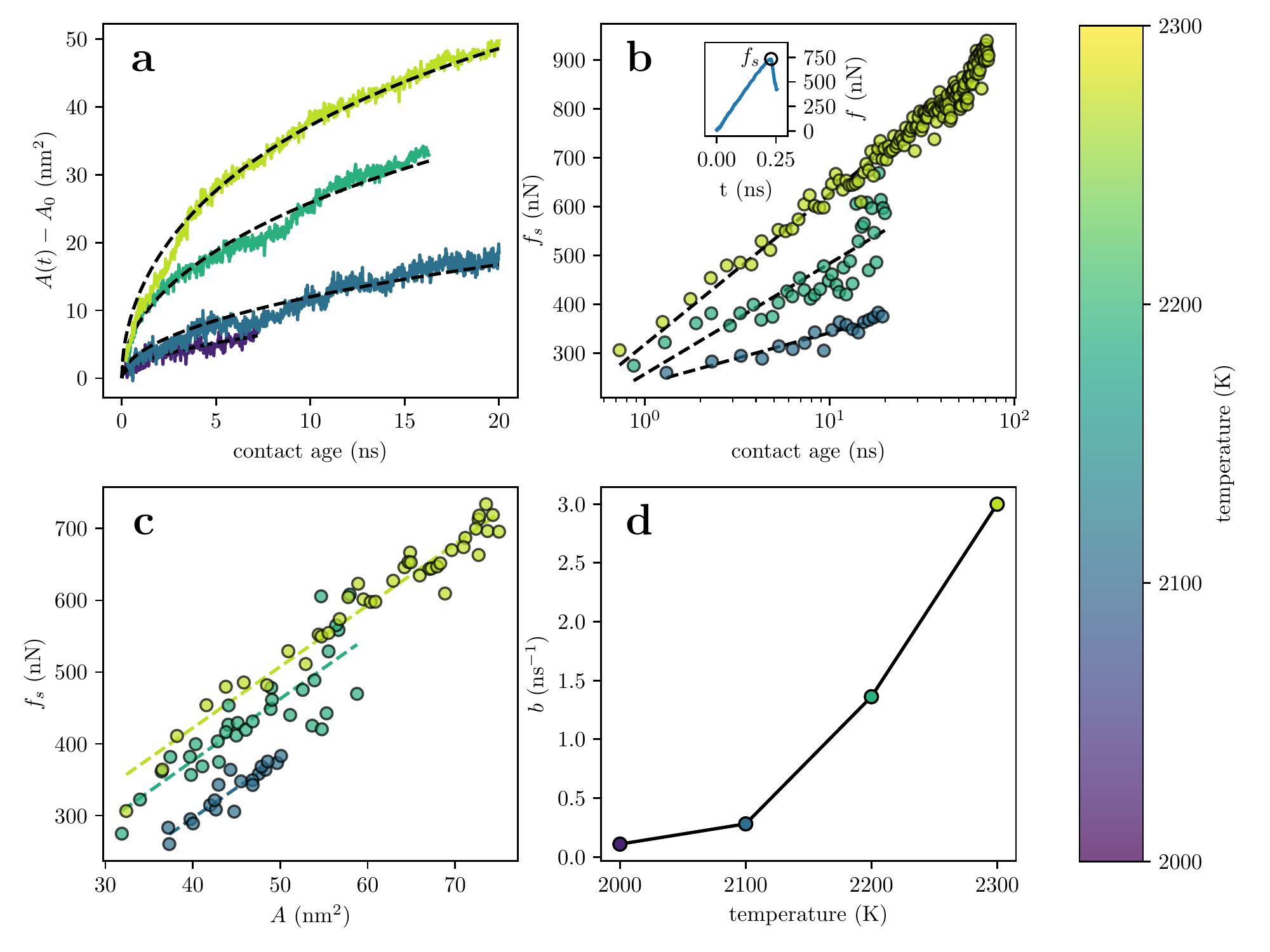}
    \caption{Contact area growth and static friction of silicon carbide nano-asperities aged without a normal load. (a) the static friction force as a function of contact age. (b) the increase in real contact area as a function of contact age, with fits to the contact area model in equation \ref{eq:contactarea} on top (dashed lines). (c) the static friction force as a function of contact area. (d) the values of the fitting parameter $b$ from the contact area model as a function of temperature.}
    \label{fig:figure3}
\end{figure}

\begin{figure}
    \centering
    \includegraphics[width=11.8cm]{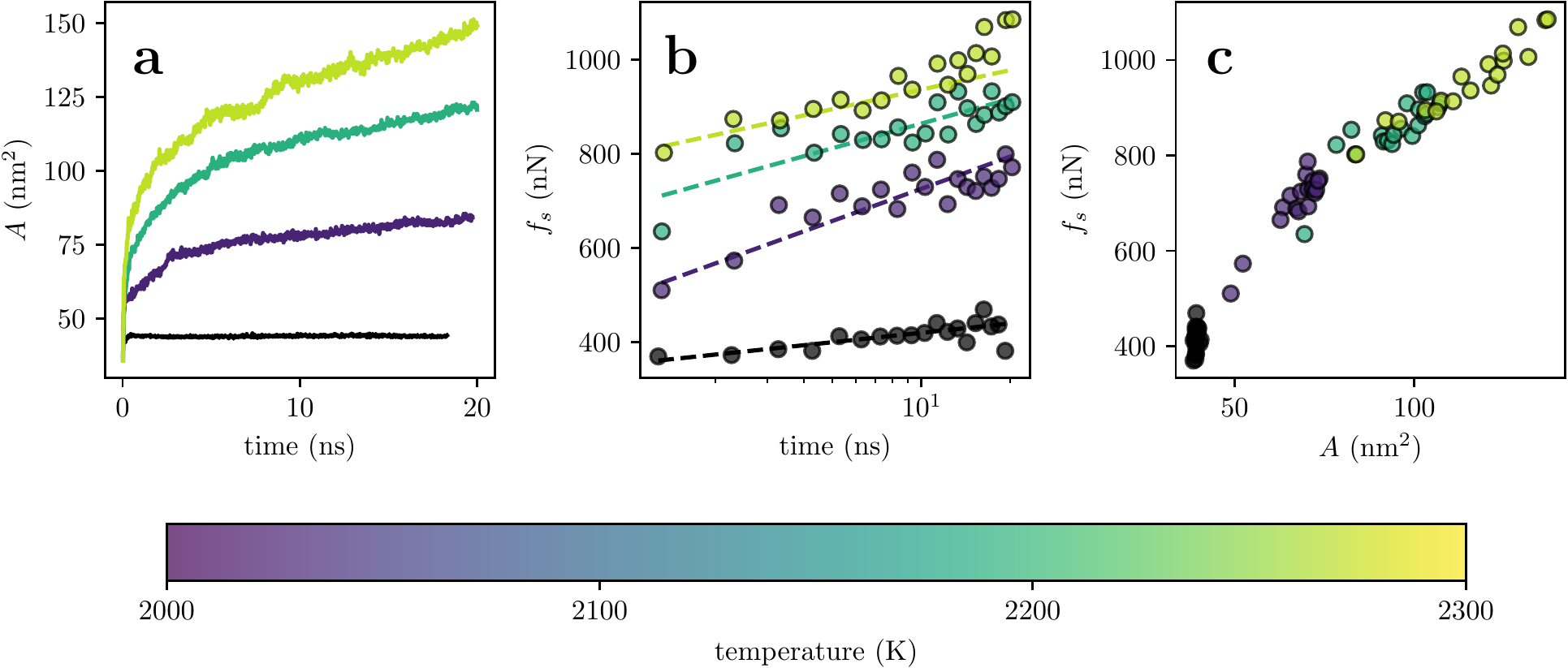}
    \caption{Contact area growth and static friction of silicon carbide nano-asperities aged under a normal load of $P=480$ MPa. (a) the static friction force as a function of contact age. (b) real contact area as a function of contact age. (c) the static friction force as a function of contact area.}
    \label{fig:figure4}
\end{figure}

\begin{figure*}
    \centering
    \includegraphics[width=12cm]{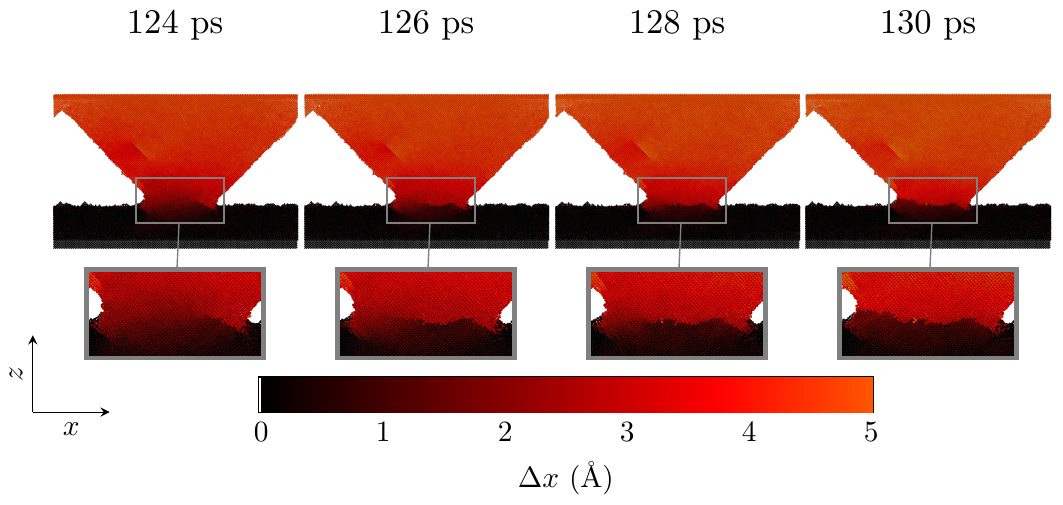}
    \caption{Cross section view of an asperity undergoing rupture for four different times given relative to the start of deformation. Atoms colors indicate their displacement in $x$-direction ($\Delta x$) since the onset of deformation. The rupture area is magnified and displayed in separate windows. The simulation was performed at a temperature of $T=2000$ K, normal pressure of $P=480$ MPa, and the system was relaxed for 18 ns before deformation.}
    \label{fig:crack}
\end{figure*}

The contact area as a function of contact age is shown in figures \ref{fig:figure3}a (no normal load) and \ref{fig:figure4}a (480 MPa normal load).
The contact area grows with time in all examined scenarios, and it grows faster with higher temperature and with a higher normal load.
A contact area model for frictional aging due to pure diffusion, as presented in section \ref{sec:model}, is fitted to the contact area in figure \ref{fig:figure3}.
As mentioned in the section \ref{sec:method}, we take configurations from multiple different contact ages and measure the maximum static friction that the system can sustain. We take the static friction as the maximum recorded shear force in the shear loading simulations, as shown in the inset of figure \ref{fig:figure3}b.
We recall that static friction is always measured at 1200 K, while the temperature indicates the system temperature under relaxation.
Not surprising, the static friction exhibits a logarithmic trend with time, as shown in figures \ref{fig:figure3}b (no normal load) and \ref{fig:figure4}b (480 MPa normal load).
The relationships between contact age and static friction force for various temperatures ($T\in{2000, 2100, 2200, 2300}$ K) are shown in figure \ref{fig:figure3}c (no normal load) and \ref{fig:figure4}c (480 MPa normal load). 
For the case with no normal load, the friction-area curves all display a linear behavior, but they are shifted with respect to each other and do not overlap.
For the case with a 480 MPa normal load, the friction-area curves clearly have different slopes, and do either not have a linear behavior or do not overlap for the entire range of areas. The relationship between friction force and contact area is discussed in section \ref{sec:frictionarea}.
Lastly, the typical asperity rupture under shear deformation is illustrated in figure \ref{fig:crack}.
We observe that ruptures propagate either from the leading edge to the trailing notch or in the opposite direction, which is debated in section \ref{sec:failure}.

\subsection{Model for the contact area evolution} \label{sec:model}
Visualizing the evolution of the asperity throughout the simulation reveals that more atoms enter than leave the notch region (that is, the vacuum in the notch between the substrate and the asperity).
This is the reason behind the increase in contact area.
By color-coding the atoms according to their displacement, we observe that most atoms originate from the substrate.
Since there is no external force acting on the atoms, the process is governed solely by diffusion.
Our observations suggest that the probability of entering the region is higher than leaving it, possibly due to a decrease in surface free energy when the notch region is filled with atoms.
According to nucleation theory, atoms need to overcome an energy barrier to transition between stable and metastable states. The rate of this process depends on temperature through the Arrhenius law, which relates the rate constant $\kappa$ to the activation energy $E_a$ and the attempt frequency $A$ as $\kappa=A\exp(-E_a/k_BT)$. When the energy barrier or attempt frequency required to enter the region is lower than that needed to exit it, atoms tend to accumulate in the region.
Moreover, as the notch region becomes filled up, there is a gradual decrease in surface free energy, which causes the rate of increase in contact area to decrease over time.
This phenomenon can be modeled by a time-increasing energy barrier, as proposed in Ref. \cite{brechet1994}.
Drawing on the aforementioned reasoning, and following the methodology outlined in previous studies such as Ref. \cite{mullins2000}, we can formulate a differential equation that describes the evolution of the notch population:
\begin{subequations}
\label{eq:notchpop}
\begin{equation}
\frac{dN}{dt}=\exp\left(-\frac{Na}{2k_BT}\right) \times b(T).
\label{eq:Ndiff}
\end{equation}
Here, $N$ is the number of atoms in the notch region, $a$ is the rate of the effective energy barrier, $k_B$ is the Boltzmann constant and $b(T)$ is a temperature dependent parameter given by
\begin{equation}
    b(T)=f_1\exp\left(-\frac{E_1}{k_BT}\right) - f_2\exp\left(-\frac{E_2}{k_BT}\right),
    \label{eq:b}
\end{equation}
\end{subequations}
and has units inverse time.
The two terms represent the initial probability of an atom entering and leaving the notch region, respectively.
$E_1$ and $E_2$ are the activation energies, and $f_1$ and $f_2$ are the attempt frequencies.
We make the additional assumption that the notch population is the only time-dependent quantity.
Under this assumption, the solution of equation \ref{eq:Ndiff} is
\begin{equation}
    N(t)=N_0+\frac{k_BT}{a}\log\left(1+b(T)\frac{at}{k_BT}\right),
    \label{eq:contactatoms}
\end{equation}
where $N_0$ is the initial number of atoms in the notch region.
The mathematical form of our aging model resembles that of aging due to creep, as discussed in Ref. \cite{hatano2015}.
Similarly, frictional aging due to the formation of siloxane bridges across the interface has been explained by a thermally activated process in Ref. \cite{li2011}.
It is noteworthy that several frictional aging mechanisms with seemingly distinct causes, such as aging due to creep \cite{dieterich1994}, stronger interfacial bonds \cite{li2011}, and diffusion-driven aging (our work), can be described by thermally activated processes.
We speculate that this universal logarithmic behavior of frictional aging, as stressed by Ref. \cite{baumberger2006}, originates from thermally activated bond formation with an increasing energy barrier.

Expressing the growth of contact area in terms of the population growth in the notch region is a straightforward geometric exercise in our study. In our system, we can assume that the asperity takes the form of an inverted cone. The population in the notch region then forms a torus that is isosceles and right triangular in the poloidal direction, with a volume $V=(2\pi R)\times (l^2/4)$. Here, $R$ is the toroidal radius, and $l$ is the length of the triangular ground line. The ground area of the torus, which corresponds to the change in contact area of the asperity, is $A=4\pi lR$, and therefore $V\propto A^2$. Assuming a uniform particle density in the notch region, the torus volume is proportional to the notch population, and hence $\Delta A\propto \sqrt{\Delta N}$. For our system, the evolution of the contact area can be represented by the following equation:
\begin{equation}
A_{\text{diff}}(t)=A_0+c\left[\frac{k_BT}{a}\log\left(1+b(T)\frac{at}{k_BT}\right)\right]^{1/2},
\label{eq:contactarea}
\end{equation}
where $c$ is the constant of proportionality between the change of contact area and the change of notch population in area units.
However, the relationship between population growth and contact area may differ for a different asperity shape, although the relation $V\propto A^2$ used in equation \ref{eq:contactarea} is valid for many geometrical shapes.
Investigating diffusion-driven frictional aging with other silicon carbide asperity shapes is challenging since the asperity easily deforms into its equilibrium shape at temperatures where surface diffusion is substantial.
Freezing the asperity might be a potential solution to this issue.

The raw contact area in figure \ref{fig:figure3}b is fitted with the contact area model in equation \ref{eq:contactarea}, which has two parameters to fit: the global parameter $a$, which was determined to be $a=6.8$ eV, and the temperature-dependent parameter $b$.
Pure area fittings can disregard the proportionality parameter $c$ since it does not offer additional degrees of freedom beyond the already included parameters $a$ and $b$.
The strictly increasing behavior of $b$ is observed, as shown in figure \ref{fig:figure3}d and predicted by equation \ref{eq:b}. 
However, due to the presence of multiple variational parameters in the expression of $b$, it is not feasible to estimate the activation energies and attempt frequencies from a single fit. 
Additional simulations are required where atoms are allowed to cross the energy barrier in only one direction to estimate activation energies and attempt frequencies. 
Such simulations are outside the scope of this study.

\subsection{Friction force and contact area} \label{sec:frictionarea}
The static friction in figure \ref{fig:figure3}b shows a logarithmic increase with time at all examined temperatures, but with different slopes and intersection points with the $y$-axis.
Since the initial contact area of all the various temperatures is roughly the same, we assume that the initial contact strength is determined by the strength of individual bonds.
The slope of the logarithmic increase is mostly governed by the contact area, and it increases with temperature.
At very short times (shorter than 1 ns), we would expect to get intersecting force curves because of the competition between bond strength and contact area.
Figure \ref{fig:figure3}c shows that the friction force has a strong correlation with the contact area, which appears to be roughly proportional when there is no normal force.
While there may be a slight curvature in the data points, especially at $T=2300$ K, it is difficult to say whether this is a coincidence or a genuine effect due to the limited area range. 
Additionally, for the same contact area, the contact strengths are sorted in a way that suggests a stronger contact at higher temperatures.
Here, it is important to keep in mind that the displayed temperature is the relaxation temperature; all deformation simulations are conducted at low temperature (1200 K).
This behavior is nontrivial, and requires more investigation.
Perhaps, the bond density is lower for higher temperatures.

\subsection{Effect of a normal load}
In figure \ref{fig:figure4}a, we demonstrate that applying a normal load exhibiting a pressure of $P=480$ MPa significantly increases the initial static friction force for all temperatures.
However, this does not amplify the aging effect that leads to an increase in frictional strength over time. 
Surprisingly, the aging effect appears to be slightly weaker with the application of the normal pressure.
For example, at temperature $T=2300$ K the change in static friction force during the first 20 nanoseconds is about $\Delta f_s\approx300$ nN with a normal pressure of $P=480$ MPa, as compared to $\Delta f_s\approx400$ nN without any normal pressure.
To explain this, we have to look at the initial contact area, which was assumed to be constant in equations \ref{eq:contactatoms} and \ref{eq:contactarea}.
When a normal force is applied, the contact area increases immediately after the asperity and substrate come into contact. This leads to an increase in the toroidal radius, $R$, and thus more particles are required to increase the contact area than when $R$ is smaller.
If we look at aging due to diffusion only, the parameter $c$ in equation \ref{eq:contactarea} decreases with $R$ (and thus normal pressure).
To clarify this observation, we must examine the initial contact area, which was assumed to be constant in equations \ref{eq:contactatoms} and \ref{eq:contactarea}.
When a normal force is applied, the contact area grows immediately after the asperity and substrate make contact.
This leads to an increase in the toroidal radius, $R$, requiring more particles to grow the contact area compared to when $R$ is smaller.
If we consider aging due to diffusion alone, the parameter $c$ in equation \ref{eq:contactarea} decreases with increasing $R$ (and therefore normal pressure).
Despite the decreased aging effect, the contact area seems to increase more quickly than without a normal force (as shown in figure \ref{fig:figure4}b). 
Consequently, the static friction force is not completely proportional to the contact area.
At larger contact areas, the force per area is smaller than at smaller contact areas (as depicted in figure \ref{fig:figure4}c).
This non-proportional relationship between friction and area has been observed in several other studies, such as in Refs. \cite{dieterich1994,kilgore2012,kilgore2017,farain2022}.
Kilgore et al. suggest that any newly formed contact area lacks shear strength \cite{kilgore2012}, causing the contact area to grow faster than the friction force. 
More investigations is needed to understand whether or not this is the mechanism at work here.

In experiments conducted under normal pressures, the aging process is likely to be the result of a combination of diffusion and creep. Bréchet and Estrin developed a model to describe how the contact area evolves due to plastic creep \cite{brechet1994,hatano2015}:
\begin{equation}
A_{\text{creep}}(t)=A_0+\frac{k_BT}{P\Omega}\log\left[1+tf\frac{P\Omega}{k_BT}\exp\left(-\frac{E_a-P\Omega}{k_BT}\right)\right],
\end{equation}
where $P$ is the normal pressure, $\Omega$ is the activation volume, $f$ is the attempt frequency, and other symbols are defined as explained above. This expression is similar to our contact area model presented in equation \ref{eq:contactarea}, but there is only one activation energy since the process only goes in one direction (the probability of reverting the deformation is absent). It is also important to note that the second term in the equation does not have an exponential factor.

\begin{table}
    \centering
    \begin{tabular}{rrrrrrr}
    \hline \\
    $T$ (K) & $\rho$ (kg/m$^3$) & $c_{11}$ (GPa) & $c_{13}$ (GPa) & $c_{33}$ (GPa) & $c_{55}$ (GPa) & $c_R$ (m/s) \\\\ \hline \\
           1 & 3217 & 396.45 & 150.58 & 396.38 &  68.18 & 4494 \\
        1800 & 3038 & 286.22 & 107.68 & 297.10 &  51.83 & 4027 \\
        2000 & 3016 & 266.80 &  97.32 & 286.35 &  49.99 & 3968 \\
        2200 & 2986 & 239.72 &  70.07 & 240.47 &  46.84 & 3855 \\ \hline
    \end{tabular}
    \caption{Rayleigh wave speed and elastic constants for 3C-SiC at various temperatures. See the main text for more details.}
    \label{tab:rayleigh}
\end{table}

\subsection{Mechanical failure mechanism} \label{sec:failure}
Figure \ref{fig:crack} illustrates the shear displacement of particles in a crystalline system at 2000 K in the direction of movement as the asperity slips.
The zoomed-in inset depicts that the rupture propagates from the leading edge (end of asperity on the right-hand side) to the trailing notch (the notch on the left-hand side), which is a commonly observed rupture front.
However, there is no particular tendency in terms of which direction the crack propagates.
From the figure, the rupture velocity can be estimated to be about 1300 m/s, given that the contact's diameter is approximately 8 nm.
We employed the procedure presented in Refs. \cite{guren2022,vinh2005} to determine the Rayleigh wave speed in 3C-SiC and determine whether this rupture can be considered rapid.
For a comparable temperature, the Rayleigh wave speed in the modelled silicon carbide crystal was calculated to be 3968 m/s, which is roughly three times faster than the rupture.
Thus, the observed ruptures are relatively rapid.
Table \ref{tab:rayleigh} provides the Rayleigh wave speed of crystalline 3C-SiC at other relevant temperatures, along with the crystal density and elastic constants utilized to estimate the speed.
These constants can also be used to determine material properties like the bulk, shear and Young's modulus, as well as the Poisson ratio.
The elastic constants are consistent with the ones found in \cite{kunc1975}, and the measured Rayleigh wave speeds agree with theoretical \cite{tarasenko2021}. 

\section{Conclusions and perspectives} \label{sec:conclusions}
The aim of this study was to investigate frictional aging in a silicon carbide system, which revealed that the static friction force increases logarithmically with contact age even in the absence of a normal force.
This was attributed to the growth of the real contact area due to the notably high surface diffusion in silicon carbide. 
We developed a simple model of contact area growth based on nucleation theory and speculated that other commonly observed frictional aging mechanisms exhibit logarithmic behavior for the same reason - because they are governed by thermally activated processes with a time-increasing energy barrier.

When a normal force is applied, the static friction force is strengthened, but the aging effect weakens.
Moreover, there is a clear non-linearity between friction force and contact area in this case, which may be explained by a non-uniform stress concentration across the contact interface. 
The study found that only slow ruptures occurred, either propagating from the leading edge to the trailing notch or in the opposite direction.

In future studies, it is suggested that the contact area model proposed in equation \ref{eq:contactarea} should be validated, particularly in experiments where particles are only able to enter or leave the notch region. 
This would allow for the determination of activation energies and attempt frequencies. 
Additionally, further investigation into the non-linearity between friction and contact area should be conducted, with particular attention paid to studying interfacial stress concentration in more detail. 
Given its unique properties, silicon carbide is a material of great interest and computer simulations of it may provide valuable insights into other tribological and tribochemical phenomena. 
In a follow-up paper, we conduct sliding simulations to investigate dynamic friction, wear, and history-dependent effects.

\section{Computational details}
Molecular dynamics simulations of the nano-asperity system are performed using LAMMPS \cite{plimpton1995}, with the silicon carbide force field and parameters proposed by Vashishta et al. \cite{vashishta2007}.
To integrate the rigid body, a symplectic integrator \cite{kamberaj2005} is used and atoms in contact with the rigid body are controlled using a Nosé-Hoover thermostat \cite{nose1984,hoover1985}.
For atoms in the mobile region ($z\in[2,18.6]$ nm), the Verlet integration scheme \cite{verlet1967} is used with a time step of 1 fs for simulations with normal pressure and 2 fs for simulations without normal pressure.
A Langevin thermostat \cite{schneider1978} is applied to the regions defined by $z\in[1, 2]$ nm and [18.6, 19.6] nm with a damping time of 1.0 ps to set the temperature of the system.
Periodic boundary conditions are applied in the horizontal directions, while fixed boundaries are used in the $z$-direction.
All simulations are carried out on NVIDIA A100 graphics cards using the KOKKOS package \cite{trott2022} in LAMMPS.
Simulating the system for 1 ns takes approximately 8 hours of wall clock time with a timestep of 2 fs.
The contact area, coordination analysis, and system visualization are performed using OVITO \cite{stukowski2010}.
The presented results have been obtained using about 5,000 GPU hours.

\subsection{Friction force estimation}
The initial peak in the lateral force curve, as shown in figure \ref{fig:figure3}a, is used to measure the static friction force. 
The lateral force represents the net force acting on the upper rigid plate in the direction of its motion. 
Since the total force on the asperity becomes negative when it is moved in the positive direction, we reverse the sign of the lateral force to obtain the magnitude of the static friction force, which is of primary interest.

\subsection{Contact area measurement}
The estimation of the contact area is achieved by enveloping a surface mesh around a thin slice of the asperity located in close proximity to the surface.
This slice is determined using a combination of cluster and coordination analysis carried out in OVITO \cite{stukowski2010}.
In this context, the term ``contact area'' refers to the actual contact area as described by Mo et al. \cite{mo2009}.
More information about the procedure for measuring the contact area can be found in the supplementary material.

\subsection{Rayleigh wave speed measurement}
The speed of Rayleigh waves in 3C-SiC crystal was determined by performing a series of deformation simulations, as explained in Refs. \cite{guren2022,vinh2005}.
The simulations were carried out for various temperatures, specifically $T\in\{1, 1800, 2000, 2200, 2300\}$, using a 3C-SiC crystal with dimensions of $10\times10\times10$ nm.
To attain equilibrium, the crystal was initially relaxed in the isobaric-isothermal ensemble at $P=1$ bar for 20 ps, and then gradually adjusted to the average size during the previous simulation over 20 ps, from which we obtained the equilibrium size and density of the crystal.
Finally, the crystal was relaxed in the isothermal ensemble for another 20 ps.

To estimate the elastic constants $c_{11}$, $c_{13}$, $c_{33}$, and $c_{55}$, separate simulations were performed in which the crystal was compressed 0.1 nm in the $x$- and $z$-directions to determine the normal stresses $\sigma_{xx}$ and $\sigma_{zz}$, respectively.
The remaining elastic constant, $c_{55}$, was estimated by conducting a shear deformation simulation in the $xz$-direction to obtain the shear stress component $\sigma_{xz}$.

\backmatter

\bmhead{Supplementary information}
Supplementary information can be found here:

\bmhead{Acknowledgments}
This work is carried out under financial support by the Norwegian Research Council under grant 28704.

\bibliography{main}

\end{document}


\maketitle

\section{Introduction}
The supplementary material contains XX.

\begin{figure}[H]
    \centering
    \includegraphics[width=14cm]{fig/area_measurement/contact_marked.png}
    \caption{Caption}
    \label{fig:contact_marked}
\end{figure}

\section{Area measurement}
The area was measured by analyzing a slice of the asperity close to the substrate. The slice was chosen such that it was possible to distinguish the asperity itself from structures and peaks on the surface. We use a slice at $z=55$ \siA\, with thickness 5 \siA. The slice is marked in Figure \ref{fig:contact_marked}.

When slicing the system close to the substrate, surface structure peaks will be included in the slice. To extract the asperity only, cluster analysis is performed and only the largest cluster in the slice is kept. In figure \ref{fig:slice}, the slice with cluster analysis in illustrated before and after all small clusters are removed. The cluster radius and surface mesh probe radius have to be tuned carefully such that the asperity is correctly found as one cluster and the surface wrap reflects the actual contact area. The cluster cutoff and the probe radius is 2.7 \siA\, and 3.3 \siA, respectively.

Sometimes, the contact area is overestimated when this method is used, as peaks on the substrate connect to the asperity without contributing to the contact area. One way to avoid this is to also include a cluster analysis. We require all asperity atoms to have a coordination number of 4 within a cutoff of 2.5 \siA.

\begin{figure} [H]
    \centering
    \subfloat[Before removing clusters]{\includegraphics[width=7cm]{fig/area_measurement/area_with_clusters_lower_probe.png}}
    \subfloat[After removing clusters]{\includegraphics[width=7cm]{fig/area_measurement/area_without_clusters_lower_probe.png}}
    \caption{Obtaining the contact area slice}
    \label{fig:slice}
\end{figure}

\section{Defects formation under deformation}
Before deformation, the asperity and the substrate form a common crystal when their Miller index is aligned. What we often see under deformation, is that system first cracks up on the left-hand side, like seen by ... We also see defects across the thinnest part of the asperity.

\begin{figure}[H]
    \centering
    \subfloat[Before deformation]{\includegraphics[width=7cm]{fig/identify_diamond_structure/diamond_structure_0.png}}
    \subfloat[During deformation]{\includegraphics[width=7cm]{fig/identify_diamond_structure/diamond_structure.png}}
    \caption{Diamond structure under deformation}
    \label{fig:diamond_structure_deformation}
\end{figure}

\section{Crystals not aligned}
When the asperity and the substrate Miller indices are not aligned, they do not form a common crystal

\begin{figure}
    \centering
    \caption{Caption}
    \label{fig:my_label}
\end{figure}

\section{Crystal stability}
The asperity was carved out to be close to equilibrium, according to \cite{sveinsson2020}. In figure \ref{fig:relaxed_upper}, the shape of the asperity is visualized after 50 \sins\, relaxation at various temperatures. For temperatures 1800, 2000 and 2200 \siK\, the asperity keeps its initial shape with minor changes. The changes are more significant for temperature 2300 \siK. When looking at the diamond structure identification for the relaxed asperity, it is apparent that the initial crystal structure (3C-SiC) is still intact.

We have already seen that the surface with Miller index (110) pointing in $z$-direction is rather unstable 

\begin{figure}
    \centering
    \subfloat[$T=1800$ \siK]{\includegraphics[width=7cm]{fig/relaxed_upper/1800.png}}
    \subfloat[$T=2000$ \siK]{\includegraphics[width=7cm]{fig/relaxed_upper/2000.png}}\\
    \subfloat[$T=2200$ \siK]{\includegraphics[width=7cm]{fig/relaxed_upper/2200.png}}
    \subfloat[$T=2300$ \siK]{\includegraphics[width=7cm]{fig/relaxed_upper/2300.png}}
    \caption{Asperity shapes after 50 \sins\, of relaxation at different temperatures}
    \label{fig:relaxed_upper}
\end{figure}

\captionsetup[subfigure]{position=top}
\begin{figure}
    \centering
    \subfloat[$T=1800$ \siK]{\includegraphics[width=14cm]{fig/relaxed_lower/1800_ids_crop.png}}\\
    \subfloat[$T=2000$ \siK]{\includegraphics[width=14cm]{fig/relaxed_lower/2000_ids_crop.png}}\\
    \subfloat[$T=2200$ \siK]{\includegraphics[width=14cm]{fig/relaxed_lower/2200_ids_crop.png}}\\
    \subfloat[$T=2300$ \siK]{\includegraphics[width=14cm]{fig/relaxed_lower/2300_ids.png}}
    \caption{Relaxed lower surfaces with }
    \label{fig:my_label}
\end{figure}

\section{Rayleigh wave speed in silicon carbide}
We measure the Rayleigh wave speed in silicon carbide through the expression provided by Vinh ... through Guren:
\begin{equation}
    \frac{\rho C_R}{c_{55}}=
\end{equation}

https://www.ioffe.ru/SVA/NSM/Semicond/SiC/mechanic.html